\documentclass[a4paper,11pt]{article}
\usepackage{jheppub}
\usepackage[T1]{fontenc}
\usepackage{graphicx}
\usepackage{amsmath}
\usepackage{amssymb}
\usepackage{amsfonts}
\usepackage{bm}
\usepackage{color}

\def\be{\begin{equation}}
\def\ee{\end{equation}}
\def\bea{\begin{eqnarray}}
\def\eea{\end{eqnarray}}
\def\tr{\mathrm{tr}\, }

\def\ads{AdS$_4$ }
\def\fX{f_{_X}}
\def\fR{f_{_R}}
\def\fXX{f_{_{XX}}}
\def\fRR{f_{_{RR}}}
\def\fXR{f_{_{XR}}}
\newcommand{\mc}[1]{\mathcal{#1}}
\newcommand{\f}[2]{\frac{#1}{#2}}

\title{\boldmath Modified gravity one-loop partition function}

\author[~]{Shahab Shahidi,${}^\mu$}
\author[~]{Farid Charmchi,${}^\nu$}
\author[~]{Zahra Haghani,${}^\mu$}
\author[~]{and Leila Shahkarami${}^\mu$}

\affiliation[~]{${}^\mu$School of Physics, Damghan University, Damghan, 
	41167-36716, Iran}
\affiliation[~]{${}^\nu$School of Particles and Accelerators, Institute for 
Research in Fundamental Sciences (IPM), P.O. Box 19395-5531, Tehran, Iran}

\emailAdd{s.shahidi@du.ac.ir}
\emailAdd{charmchi@ipm.ir}
\emailAdd{z.haghani@du.ac.ir}
\emailAdd{l.shahkarami@du.ac.ir}

\abstract{
	The one-loop partition function of the $f(R,R_{\mu\nu}R^{\mu\nu})$ gravity theory is obtained around AdS$_4$ background. After suitable choice of the gauge condition and computation of the ghost determinant, we obtain the one-loop partition function of the theory. The traced heat kernel over the thermal quotient of AdS$_4$ space is also computed and the thermal partition function is obtained for this theory. We have then consider quantum corrections to the thermodynamical quantities in some special cases.
	
	}

\begin{document}
\maketitle
\section{Introduction}
It is well-known that the Einstein's theory of relativity should be modified beyond the solar system, to fit the observational data. 
At large scales, i.e., the regime of low energies, the introduction of some dark energy component seems to be necessary to explain the accelerated expansion of the universe \cite{darkenergy}. 

The interesting fact about the Einstein's gravity is that at UV limit the theory should also be modified, since it cannot explain the microscopic physics. 
The quantization of the Einstein-Hilbert action leads to a nonrenormalizable theory which is known for decades \cite{nongrav}. 
This suggests that we cannot quantize Einstein's general relativity along the path of ordinary quantum field theory. 
One can think about the possibility of modifying the dynamics of the space-time in a nonlinear manner in such a way that the theory becomes renormalizable \cite{renor}. 
Perhaps the simplest possibility is to promote the Ricci scalar in the Einstein-Hilbert action to an arbitrary function of the Ricci scalar (the $f(R)$ theories of gravity). 
The quantum corrections to the $f(R)$ theories of gravity have been extensively investigated in the literature \cite{quantfR}.

The other possibility of modifying Einstein's gravity is to break the Lorentz symmetry of the theory and then add some spatial-derivative self-interaction terms in such a way that the resulting theory becomes  renormalizable. 
This Horava-Lifshitz quantum gravity theory \cite{horava}, has been found to be very interesting both in theoretical and phenomenological point of view \cite{horavaapp}. 
Also some attempts have been made to make the theory Lorentz-invariant \cite{lorenthorava}.

Historically, the first example of a modified theory of gravity which is renormalizable at quantum level has been introduced in  \cite{stelle} in which the action
\begin{align}
S=\int d^4x \sqrt{-g}\left(\kappa^2 R+a R^2+b R_{\mu\nu}R^{\mu\nu}\right),\nonumber
\end{align}
is proven to be renormalizable for some special choices of the parameters. 
The beta function is also obtained in this case. 
This suggests that one can promote the Ricci scalar in the Einstein's theory to an arbitrary function of $R$‌ and $R_{\mu\nu}R^{\mu\nu}$, and even $R_{\mu\nu\rho\sigma}R^{\mu\nu\rho\sigma}$ (which is not necessary in 4D due to Gauss-Bonnet theorem). 
The quantum corrections to generalized Gauss-Bonnet theory over de Sitter background is investigated in \cite{dsfR}.

In this paper, we consider the $f(R,R_{\mu\nu}R^{\mu\nu})$ gravity theory over the anti-de Sitter (AdS) background. 
The choice of this Lagrangian is interesting because it can be seen as a generalization of some interesting modified gravity theories including $f(R)$ gravity, Gauss-Bonnet theory, conformal gravity and also Stelle's original work (one should however note that the $f(G)$ modified gravity theories where $G$ is the Gauss-Bonnet term cannot in general be expressed in our formalism). Also, the AdS background is interesting specially as a part of the AdS/CFT correspondence \cite{maldacena}. 
In this context, one of the most important quantities which should be computed is the partition function. 
The partition function of the gravity theory in the AdS background is then equivalent to the partition function of its dual conformal field theory.

 In this work we compute the one-loop partition function of the $f(R,R_{\mu\nu}R^{\mu\nu})$ theory around AdS$_4$ background. 
In order to do this, we fix the gauge in such a way that the resulting gauge-fixed Lagrangian becomes diagonal in the perturbation variables. 
Such a gauge fixing results in the introduction of the third ghost as discussed in \cite{odintsov}. 
Considering the ghost determinants and the Jacobian of the field transformation, we arrive at the one-loop partition function of the theory. 
One should note that the procedure introduced in this paper is a little bit different from the one  usually done in the literature. 
We then discuss how we can recover the standard results in some special cases. 

Using the heat kernel method, we are able to evaluate the determinants appearing in the partition function and compute the traced heat kernel over the thermal quotient of the AdS$_4$ space, to obtain the thermal partition function in $f(R,R_{\mu\nu}R^{\mu\nu})$ theory. 
The traced heat kernel over thermal quotient of the AdS space in the conformal gravity theory is discussed in \cite{iva}.
We then use this method to obtain thermodynamical quantities such as Helmhotz free energy and also the entropy which should be viewed as a one-loop quantum correction to the entropy. 
We furthermore compare the consequences of the quantum corrections to the thermodynamical quantities in some special cases of interest.

We should note that while this work was in progress, the preprint \cite{akhary} appeared on arXiv in which the authors have done a similar calculation as is presented in this manuscript and obtain the one-loop divergences of the same theory. 
However, our calculations are different in the way that besides using a different approach for fixing the gauge, we are also interested in the thermodynamical implications of the theory over thermal quotient of the AdS$_4$ space which is not the scope of \cite{akhary}.
%===================================================
\section{The model}\label{sec2}
Let us consider the following general theory containing a function of both the Ricci scalar and the square of the Ricci tensor
\begin{align}\label{eq2}
S=\int d^4x\sqrt{-g}\, f(R,R_{\mu\nu}R^{\mu\nu}),
\end{align}
where $f$ is an arbitrary function with mass dimension $4$.
For our purposes, we need to calculate the second variation of this action on shell by imposing the equation of motion, coming from the first variation.
By varying once the above action with respect to the metric tensor, one may obtain
\begin{align}\label{2}
\delta S=\int d^4x\sqrt{-g}\,B_{\mu\nu}\,\delta g^{\mu\nu},
\end{align}
where $B_{\mu\nu}$ is the equation of motion which can be written as
\begin{align}\label{3}
B_{\mu\nu}=\fR R_{\mu\nu}-\f12fg_{\mu\nu}+2\fX R_{\alpha\mu}R^\alpha_{~\nu}+\Box H_{\mu\nu}+\nabla_\alpha\nabla_\beta H^{\alpha\beta}g_{\mu\nu}-2\nabla_\alpha\nabla_{(\mu}H_{\nu)}^{~\alpha}=0.
\end{align}
Here $H_{\mu\nu}=\fR\,g_{\mu\nu}+2\fX R_{\mu\nu}$ and we have defined $X\equiv R_{\mu\nu}R^{\mu\nu}$,
$\fR\equiv\partial f/\partial R$ and $\fX\equiv\partial f/\partial X$.

It can be easily seen that the above equation acquires AdS$_4$ solution. 
As we know, for the \ads space we have in general
\begin{align}
	\bar{R}_{\mu\nu\rho\sigma}=\f{\Lambda}{3}(\bar{g}_{\mu\rho}\bar{g}_{\nu\sigma}-\bar{g}_{\mu\sigma}\bar{g}_{\nu\rho}),\quad \bar{R}_{\mu\nu}=\Lambda \bar{g}_{\mu\nu},\quad \bar{R}=4\Lambda.
\end{align}
Here and in the following few equations, we use the bar sign to denote the background metric and specially the \ads background which we are most interested in %which are of most interest to us
 as a sample background in this paper.
Using the above relations, the equation of motion \eqref{3} is reduced to
\begin{align}\label{adseq}
	2\fX \Lambda^2+\fR\Lambda-\f12f=0,
\end{align}
which can be solved for $\Lambda$ if one fixes the function $f$. For example, for the simple case $f=\kappa^2 R+\eta R\, R_{\mu\nu}R^{\mu\nu}$ one has an \ads solution with $\Lambda=-\sqrt{\kappa^2/4\eta}$.

Now, we proceed to find the second variation of the action.
Consider the small fluctuation $h_{\mu\nu}$ of the tensor metric $g_{\mu\nu}$ around the background metric $\bar{g}_{\mu\nu}$ as follows
\begin{align}
g_{\mu\nu}=\bar{g}_{\mu\nu}+h_{\mu\nu}.
\end{align}
Varying the relation \eqref{2} and evaluating it on shell for the background metric $\bar{g}_{\mu\nu}$, we arrive at
\begin{align}
\delta^{(2)}S=&\int d^4x\sqrt{-\bar{g}}
\left(\bar{\delta B}^{\mu\nu}h_{\mu\nu}+\bar{B}^{\mu\nu}\delta h_{\mu\nu}-\f12\bar{g}_{\alpha\beta}\bar{B}_{\mu\nu}\,h^{\alpha\beta}h^{\mu\nu}\right)
\nonumber\\
=&\int d^4x\sqrt{-\bar{g}}\,\bar{\delta B}^{\mu\nu}h_{\mu\nu}.
\end{align}
Again tensors with the bar sign are calculated in the background metric. 
The last two terms in the first line vanish by imposing the equation of motion \eqref{3}.
Our task in the next section is to obtain the one-loop parition function for the theory of our interest.
We use the definition $\mc{L}_2\equiv\bar{\delta B}^{\mu\nu}h_{\mu\nu}$ in what follows, for brevity of notation.
We also omit the bar sign in the following sections, since all the calculations are performed in the background metric.
%===================================================
\section{The one-loop partition function of $f(R,R_{\mu\nu}R^{\mu\nu})$ gravity}\label{secpf}
Let us now compute the one-loop partition function of the gravitational theory \eqref{eq2}. The partition function at the one-loop level can be obtained from the path integral over all fluctuations around the background ($h_{\mu\nu}$) as
\begin{align}\label{eq1}
Z^{\textmd{1-loop}}=\int Dh_{\mu\nu}e^{-\delta^{(2)}S}.
\end{align}
Because our theory is generally covariant, it has a gauge freedom of the form
\begin{align}\label{eq1a}
\delta h_{\mu\nu}=\nabla_\mu\epsilon_\nu+\nabla_\nu\epsilon_\mu,
\end{align}
in which $\epsilon_\mu$ denote the parameters of the gauge group.
In order to calculate the integral \eqref{eq1} only over physical degrees of freedom among all the ones that are related to each other through the gauge transformation \eqref{eq1a}, we have to fix the above gauge freedom. 
Consider the following gauge condition
\begin{align}
\xi^\mu-l^\mu=0,
\end{align}
where $l^\mu$ is some arbitrary function and
\begin{align}
\xi_\mu=\nabla_\alpha h^\alpha_{~\mu}-\left(\rho+\f14\right)\nabla_\mu h .
\end{align}
Notice that the above condition reduces to the Lorentz condition for $\rho=\frac{1}{4}$.
To impose the gauge-fixing condition on the functional integral \eqref{eq1}, we must multiply the integrand by $\delta(\xi^\mu-l^\mu)$ (see the details in \cite{odintsov}).
This leads to the following result
\begin{align}\label{eq4}
Z^{\textmd{1-loop}}=\int Dh_{\mu\nu}e^{-\int d^4x\sqrt{-g}\mc{L}_2}\delta(\xi^\mu-l^\mu)\Delta[h],
\end{align}
where $\Delta[h]$ is related to the gauge group of the theory, which can be found by the use of the Faddeev-Popov ansatz.

Inspired by the Gaussian integral, one can write the following identity
\begin{align}\label{eq3}
1=\textmd{det}^{1/2}G_{\alpha\beta}\int Dl\,e^{-\f12\!\int\! d^4x\sqrt{-g}\left(l^\alpha G_{\alpha\beta} l^\beta\right)},
\end{align}
where $G_{\alpha\beta}$ is an arbitrary functional of $h_{\mu\nu}$. 
Multiplying \eqref{eq4} by the above unity, we arrive at the folowing expression for the one-loop partition function
\begin{align}\label{eq5}
Z^{\textmd{1-loop}}=\int Dh_{\mu\nu} \Delta[h]\textmd{det}^{1/2}G_{\alpha\beta} \,e^{-\!\int\! d^4x\sqrt{-g}(\mc{L}_2+\mc{L}_{gf})},
\end{align}
where we have defined the gauge fixing Lagrangian as
\begin{align}
\mc{L}_{gf}\equiv \f12\;\xi_\mu G^{\mu\nu}\xi_\nu,
\end{align}
and 
\begin{align}\label{gf}
G^{\mu\nu}=\alpha(-g^{\mu\nu}\Box-\gamma\nabla^\mu\nabla^\nu+\nabla^\nu\nabla^\mu)+\Lambda g^{\mu\nu},
\end{align}
where $\alpha$ and $\gamma$ are some constants and $\Lambda$ is the AdS$_4$ parameter.
Now using the Faddeev-Popov ansatz and the above calculations, one can easily calculate the form of $\Delta[h]$ as
\begin{align}
\Delta[h]=\textmd{det} M^\alpha_\beta[h],
\end{align}
with
\begin{align}
M^\alpha_\beta[h]=G^\alpha_\mu\f{\delta\xi^\mu}{\delta\epsilon^\beta}.
\end{align}
The second part of the multiplication in the right-hand side of the above equation can be calculated as
\begin{align}
\f{\delta\xi^\mu}{\delta\epsilon^\beta}=\delta^\mu_\beta\Box+R^\mu_{\beta}+\left(\f{1-4\rho}{2}\right)\nabla^\mu\nabla_\beta.
\end{align}
Substituting the corresponding relations for the determinants into \eqref{eq5}, we finally end up with the following relation for the one-loop partition function 
\begin{align}\label{oneloop}
Z^{\textmd{1-loop}}=\int Dh\, DC\, D\bar{C}\, Db\, e^{-\int d^4x\sqrt{-g}(\mc{L}_2+\mc{L}_{gf}+\mc{L}_{gh})},
\end{align}
where the ghost Lagrangian is defined as
\begin{align}
\mc{L}_{gh}=\f12\bar{C}^\alpha M_{\alpha\beta}[h] C^\beta+\f12b^\alpha G_{\alpha\beta} b^\beta.
\end{align}
In the above relation, $C$ and $b$ are complex and real Grossman variables, respectively, and we have used the relations
\begin{align}
\textmd{det}M_{\alpha\beta}=\int DC\, D\bar{C}\,e^{-\f12\!\int\! d^4x\sqrt{-g}\left(\bar{C}^\alpha M_{\alpha\beta} C^\beta\right)},
\end{align}
and
\begin{align}
\textmd{det}^{1/2}G_{\alpha\beta}=\int Db\,e^{-\f12 \!\int\! d^4x\sqrt{-g}\left(b^\alpha G_{\alpha\beta} b^\beta\right)}.
\end{align}
The anticommutating fields $C$ are called the Faddeev-Popov ghosts and the $b$ field is called the third ghost.

Let us now York decompose the metric fluctuation $h_{\mu\nu}$ into the transverse-traceless part $A_{\mu\nu}$, the trace part $h$, the helicity-0 part $\chi$ and the vector part $T$ as
\begin{align}\label{317}
h_{\mu\nu}=A_{\mu\nu}+\frac{1}{4}g_{\mu\nu}h+2\nabla_{(\mu}T_{\nu)}+2\nabla_\mu\nabla_\nu\chi-\f12 g_{\mu\nu}\nabla_\alpha\nabla^\alpha\chi,
\end{align}
where we have $\nabla^\mu A_{\mu\nu}=0=g^{\mu\nu}A_{\mu\nu}$ and $\nabla^\mu T_\mu=0$.
Also, we decompose the ghost fields as
\begin{align}
&C_\alpha=C^\perp_\alpha+\nabla_\alpha C,\\
&b_\alpha=b^\perp_\alpha+\nabla_\alpha b,
\end{align}
with $\nabla^\mu C^\perp_\mu=0=\nabla^\mu b^\perp_\mu$.

By applying the York decomposition on the second variation of the original perturbed action around the background metric, ${\cal L}_2$ reads
\begin{align}\label{act1}
\f12{\cal L}_2=A_{\mu\nu}{\cal{O}}_{_{\!AA}}A^{\mu\nu}+T_\mu{\cal{O}}_{_{\!TT}}T^\mu+\chi{\cal{O}}_{_{\!\chi\chi}}\chi+h{\cal{O}}_{_{\! hh}}h+\chi{\cal{O}}_{_{\!\chi h}}h.
\end{align}
The operators $\cal{O}$ can be expressed in terms of some powers of $\Box$. The full form of these operators are presented in Appendix \ref{app1}.

We also apply the York decomposition on the other two Lagrangians appeared in \eqref{oneloop}.
The gauge-fixing Lagrangian $\mc{L}_{gf}$ can be computed as
\begin{align}
\mc{L}_{gf}=T_{\mu}\mc{O^\prime}_{_{TT}}T^\mu+\xi\mc{O^\prime}_{_{\chi\chi}}\xi+h\mc{O^\prime}_{_{hh}}h+h\mc{O^\prime}_{_{h\chi}}\xi,
\end{align}
where
\begin{align}
&\mc{O^\prime}_{_{TT}}=-\f\alpha2\Box^3+\f12(1-\alpha)\Lambda+\f12(2+\alpha)\Lambda^2\Box+\f12(1+\alpha)\Lambda^3,\\
&\mc{O^\prime}_{_{\chi\chi}}=\f98\alpha\gamma\Box^4+\f38(8\alpha\gamma-3)\Lambda\Box^3+(2\alpha\gamma-3)\Lambda^2\Box^2-2\Lambda^3\Box,\\
&\mc{O^\prime}_{_{hh}}=\f12\alpha\rho^2\gamma\Box^2-
\f12\rho^2\Lambda\Box,\\
&\mc{O^\prime}_{_{h\chi}}=-\f32\alpha\rho\gamma\Box^3+
\f12\rho(3-4\alpha\gamma)\Lambda\Box^2+2\rho\Lambda^2\Box.
\end{align}
The ghost Lagrangian can be written as
\begin{align}
\mc{L}_{gh}=C^{\perp\mu}\mc{O}_1 C^\perp_\mu+C\mc{O}_2 C+b^{\perp\alpha}\mc{O}_3 b^\perp_\alpha+b\ \mc{O}_4 b,
\end{align}
where the operators have the forms
\begin{align}
\mc{O}_1&=-\alpha\Box^2+\Lambda\Box+(1+\alpha)\Lambda^2,\\
\mc{O}_2&=\f12(-\Box)\bigg(\alpha(4\rho-3)\gamma\Box^2+
(3-4\rho-4\alpha\gamma)\Lambda\Box+4\Lambda^2\bigg),\\
\mc{O}_3&=-\f12\bigg(\alpha\Box-(1+\alpha)\Lambda\bigg),\\
\mc{O}_4&=\f{1}{2}(-\Box)(-\alpha\gamma\Box+\Lambda).
\end{align}
%\begin{align}
%\mc{O}_1&=\f12(-\Box-\Lambda)(-\Box+2\Lambda),\\
%\mc{O}_2&=-\f12(-\Box)(-\Box+\f{4\Lambda}{\beta-3})(-\Box+\f{\Lambda}{\gamma}),\\
%\mc{O}_3&=\f12(-\Box+2\Lambda),\\
%\mc{O}_4&=\f{\gamma}{2}(-\Box)(-\Box+\f{\Lambda}{\gamma}).
%\end{align}
It can be easily checked that by assuming
\begin{align}
\alpha=1,\qquad \rho=-\f{a_1}{2\Lambda^2},\qquad  \gamma=-\f{4a_3\Lambda^2}{3a_1},
\end{align}
we obtain $\mc{O}_{h\chi}+\mc{O^\prime}_{h\chi}=0$, and therefore the action $\mc{L}_{\textmd{gauged}}\equiv\mc{L}_2+\mc{L}_{gf}$ becomes diagonal.
Here, $a_1$ and $a_3$ are two of the parameters appearing in $\mc{O}_{_{h\chi}}$ which are presented in Appendix \ref{app1}.
After some algebra, $\mc{L}_{\textmd{gauged}}$ can be simplified as 
\begin{align}
\mc{L}_{\textmd{gauged}}=A_{\mu\nu}\mc{O}^{\prime\prime}_{_{AA}}A^{\mu\nu}
+T_\mu\mc{O}^{\prime\prime}_{_{TT}}T^\mu+h\mc{O}^{\prime\prime}_{_{hh}}h+\chi\mc{O}^{\prime\prime}_{_{\chi\chi}}\chi,
\end{align}
in which
\begin{align}
\mc{O}^{\prime\prime}_{_{AA}}&=(-\Box+m^2_{A,+})(-\Box+m^2_{A,-}),\nonumber\\
\mc{O}^{\prime\prime}_{_{TT}}&=-(-\Box-\Lambda)(-\Box+m^2_{T,+})(-\Box+m^2_{T,+}),\nonumber\\
\mc{O}^{\prime\prime}_{_{hh}}&=(-\Box+m^2_{h,+})(-\Box+m^2_{h,-}),\nonumber\\
\mc{O}^{\prime\prime}_{_{\chi\chi}}&=(-\Box)(-\Box-\f{4\Lambda}{3})(-\Box+m^2_{\chi,+})(-\Box+m^2_{\chi,+}).
\end{align}
We have written the explicit expressions of the above masses in Appendix \ref{app2}.

The final form of the one-loop partition function of our model is written as
\begin{align}\label{pf}
	&Z^{\textmd{1-loop}}=Z_{\textmd{gauged}}Z_{gh}Z_{N_{h}}Z_{N_{C}}Z_{N_{b}}\nonumber\\&=
\det_{(1)}(-\Box-\Lambda)\det_{(0)}\left(-\Box+\f{4\Lambda}{\rho-3}\right)\left[\det_{(1)}(-\Box+2\Lambda)\det_{(0)}\left(-\Box+\f{\Lambda}{\gamma}\right)\right]^{3/2}\nonumber\\&\times\bigg[\det_{(2)}(-\Box+m^2_{A,+})\det_{(2)}(-\Box+m^2_{A,-})\det_{(1)}(-\Box+m^2_{T,+})\det_{(1)}(-\Box+m^2_{T,-})\nonumber\\
	&\times\det_{(0)}(-\Box+m^2_{h,+})\det_{(0)}(-\Box+m^2_{h,-})\det_{(0)}(-\Box+m^2_{\chi,+})\det_{(0)}(-\Box+m^2_{\chi,-})\bigg]^{-1/2}.
\end{align}
The last three partition functions in the first line of the above relation are associated with the decomposition of the fields which are written in Appendix \ref{app3}.

Now, let us consider some special cases of interest of the partition function \eqref{pf}.

 In the case of the Einstein gravity with the Lagrangian $\mathcal{L}=\kappa^2(R-2\Lambda)$, the appropriate choice of the gauge fixing parameters \eqref{gf} in which the scalar part of $\mathcal{L}_{\textmd{gauged}}$ becomes diagonal would be
$$\alpha=0,\qquad\rho=-\f{\kappa^2}{2\Lambda},\qquad\gamma=1,$$ 
and the one-loop partition function \eqref{pf} reduces to
\begin{align}
&Z_{\textmd{Einstein}}^{\textmd{1-loop}}=\sqrt{\f{\det_{(1)}(-\Box-\Lambda)}{{\det_{(2)}}\left(-\Box+\f{2\Lambda}{3}\right)}},
\end{align}
which is well known \cite{tseytlin}.

Also, we can recover the result of the one-loop partition function of the conformal gravity theory, first presented in \cite{iva}. 
In this case, the Lagrangian takes the form $\mathcal{L}=\eta C^\alpha_{~\beta\mu\nu}C_\alpha^{~\beta\mu\nu}$ which can be converted to a simpler form $\mathcal{L}=\f23\eta(3R_{\mu\nu}R^{\mu\nu}-R^2)$ using the Gauss-Bonnet theorem. It can be seen that the appropriate choice of the gauge-fixing parameters in this case can be obtained as
$$\alpha=0,\qquad\rho=0,\qquad\gamma=1,$$ 
and the partition function becomes
\begin{align}
&Z^{\textmd{1-loop}}=Z_{\textmd{Einstein}}^{\textmd{1-loop}}\sqrt{\f{\det_{(0)}(-\Box-\f{4\Lambda}{3})}{{\det_{(2)}}\left(-\Box+\f{4\Lambda}{3}\right)}},
\end{align}
which matches with the results of \cite{iva}, noting that they use $\eta=-2$‌ and $\Lambda=-3$.

Now, let us consider a special case of $f(R)$ gravity theory, namely $\mathcal{L}=\kappa^2 R+\eta R^3$. This theory admits an AdS background with $\Lambda=-\kappa/4\sqrt{\eta}$ provided that $\eta>0$. 
In this case, the appropriate choice of the gauge-fixing parameters would be
$$\alpha=1,\qquad\rho=\f{\kappa^2}{\Lambda},\qquad\gamma=-\f94,$$ 
and the partition function takes the form
\begin{align}
&Z^{\textmd{1-loop}}=Z_{\textmd{Einstein}}^{\textmd{1-loop}}\det_{(1)}(-\Box+2\Lambda)\sqrt{{\det_{(0)}}\left(-\Box-\f{4\Lambda}{9}\right)}.
\end{align}

At the end, for the theory $\mathcal{L}=\kappa^2R+\eta RR_{\mu\nu}R^{\mu\nu}$ discussed in section \ref{sec2}, the appropriate choice of the gauge-fixing parameters is
$$\alpha=1,\qquad\rho=\kappa^2,\qquad\gamma=-\f52,$$ 
and the partition function becomes
\begin{align}\label{pfads}
	Z^{\textmd{1-loop}}=Z_{\textmd{Einstein}}^{\textmd{1-loop}}\det_{(1)}(-\Box+2 \Lambda)\sqrt{\f{\det_{(0)}\left(-\Box-\frac{2\Lambda}{5}\right)}{\det_{(2)}\left(-\Box-\frac{10\Lambda}{3}\right)}}.
\end{align}
From the above examples, one can deduce that the $f(R)$ gravity theories produce only one spin-2 contribution to the partition function, which is related to the massless graviton of the Einstein's theory. 
However, adding terms containing $R_{\mu\nu}R^{\mu\nu}$ produces a secondary tensor contribution to the partition function.
%===================================================
\section{The traced heat kernel on AdS$_4$}
This section is devoted to a short explanation of the heat kernel method and the prescription of \cite{iva}, used here to compute the determinants appearing in the obtained one-loop partition functions of the previous section.

We can define the heat kernel of the sample operator $\Delta_{(s)}$, for a field with the spin $s$, on a manifold ${\cal M}$ between two points $x$ and $y$ as
\begin{align}
K_{ab}^{(s)}\left(x,y;t\right)=\left\langle y,b\left| e^{t\Delta_{(s)}}\right|x,a\right \rangle,
\end{align}
where $a$ and $b$ denote Lorentz indices of the field.
The partition function related to the trace of the heat kernel over both the spin indices and space-time is written as
\begin{align}
\ln Z_{(s)}=-\frac{1}{2}\ln \det \left(-\Delta_{(s)}\right)=-\frac{1}{2}\tr \ln \left(-\Delta_{(s)}\right)=-\frac{1}{2}\int_{0}^{\infty}\frac{dt}{t}K^{(s)}(t),
\end{align}
in which the trace of the heat kernel is defined as
\begin{align}
K^{(s)}(t)\equiv \tr e^{t\Delta_{(s)}}=\int_{{\cal M}}
d^d x\sqrt{g}  \sum_a K_{aa}^{(s)}(x,x;t),
\end{align}
Although the computation of the heat kernel in general manifolds is a difficult task even for the scalar operators, for symmetric spaces the use of the harmonic analysis on group manifolds can help us to make the problem tractable.
For example for the trace of the heat kernel on the thermal quotient $\textmd{AdS}_4$, the quotient space can be assummed as $H^4\simeq SO(4,1)/SO(4)$.
For this case the trace of the heat kernel becomes
\begin{align}
K^{(s)}(t)=\frac{\beta}{2 \pi} \sum_{k\in \mathbb{Z}}\sum_{\vec{m}}\int_0^{\infty}d\lambda
 \chi_{_{\lambda,\vec{m}}}\left(\gamma^{k}\right)e^{tE_R^{(s)}},
\end{align}
where $E_R^{(s)}$s are the eigenvalues of the spin-s operator on the quotient space $H^4$, $\chi_{_{\lambda,\vec{m}}}\left(\gamma^{k}\right)$ is the Harish-Chandra in principle series $SO(4,1)$, $\gamma$ is an element of the thermal quotient of $S^4$, $\beta$ is the inverse temperature ($\beta=\frac{1}{kT}$) and $(\lambda, \vec{m})$ denotes the principal series representation \cite{iva}.

Now, we need to obtain the eigenvalues for the symmetric transverse traceless tensor that we are interested in.
The unitary irreducible representations of $SO(4,1)$ and $SO(4)$ denoted by $R$ and $S$, repectively, are characterised by the following array
\begin{align}
R=&\left(m_1,m_2\right)=\left(i\lambda,m_2 \right)~~\mathrm{with}~\lambda\in \mathbb{R},\nonumber\\
S=&\left(s_1,s_2\right)~~~~~~~~~~~~~~~~~~~\mathrm{with}~s_1\geqslant s_2\geqslant 0 ,
\end{align}
in which $m_2$ is a non-negative (half-) integer, and $s_1$ and $s_2$ are (half-) integer.
The eigenvalues of the spin-s operator in the quotient space $H^4$ are given by
\begin{align}\label{energy}
-E_R^{(s)}=C_2(R)-C_2(S).
\end{align}
Here, $C_2(R)$ and $C_2(S)$ are the Casimirs of the unitary irreducible representation of $SO(4,1)$ and $SO(4)$, respectively, which can be written as
\begin{align}
C_2(R)&=m^2+2r^{SO(4,1)}\cdot m,\nonumber\\
C_2(S)&=s^2+2r^{SO(4)}\cdot s,
\end{align}
where the dot sign denotes the usual Euclidean product, and $$r^{SO(4,1)}_i=\frac{5}{2}-i,\qquad r^{SO(4)}_i=2-i,\qquad \textmd{for}\quad i=1,2,3.$$
Using these relations for the Casimirs, the equation \eqref{energy} is obtained as
\begin{align}
-E_R^{(s)}=\lambda^2+\frac{9}{4}+S,
\end{align}
In the principal series of $SO(4,1)$, the Harish-Chandra character is written as follows
\begin{align}
\chi_{_{\lambda,\vec{m}}}\left(\beta,\phi_1\right)=\frac{\left(e^{-i\beta\lambda}+e^{i\beta\lambda}\right)\chi_{\vec{m}}^{SO(3)}\left(\phi_1\right)}{e^{-3\beta/2}\left|e^{\beta}-1\right|\left|e^{\beta}-e^{i\phi_1}\right|^2},
\end{align}
in which $\chi_{\vec{m}}^{SO(3)}$ is the character of the $SO(3)$ representation.
For the thermal quotient we have $\phi_1=0$ and $\chi_{\vec{m}}^{SO(3)}(0)=1+2S$.
Using these values, we can find the traced heat kernel as
\begin{align}
K^{(s)}(t)=\frac{\beta (1+2 S)}{8\sqrt{\pi t}}\sum_{k\in \mathbb{Z}_+}\frac{e^{-\frac{k^2\beta^2}{4t}-t\left(\lambda^2+S\right)}}{\sinh^3 \frac{k\beta}{2}}.
\end{align}
To prevent the divergence of this equation, we have excluded $k=0$.
Finally, we can evaluate the partition function for the arbitrary spin as follows
\begin{align}\label{pf1}
\ln Z_{(s)}=-(1+2 S)\sum_{k\in \mathbb{Z}_+}\frac{e^{-k\beta\left(\frac{3}{2}+\sqrt{\frac{9}{4}+m^2+S} \right)}}{\left(1-e^{-k\beta}\right)^3k}.
\end{align}

Using the above relation, we can simply complete the calculation of the one-loop partition function for the cases of our interest with the \ads background.
The results are presented in the following section.
%===================================================
\section{Thermodynamical implications}\label{thermo}
\begin{figure}
	\centering
	\includegraphics[scale=0.35]{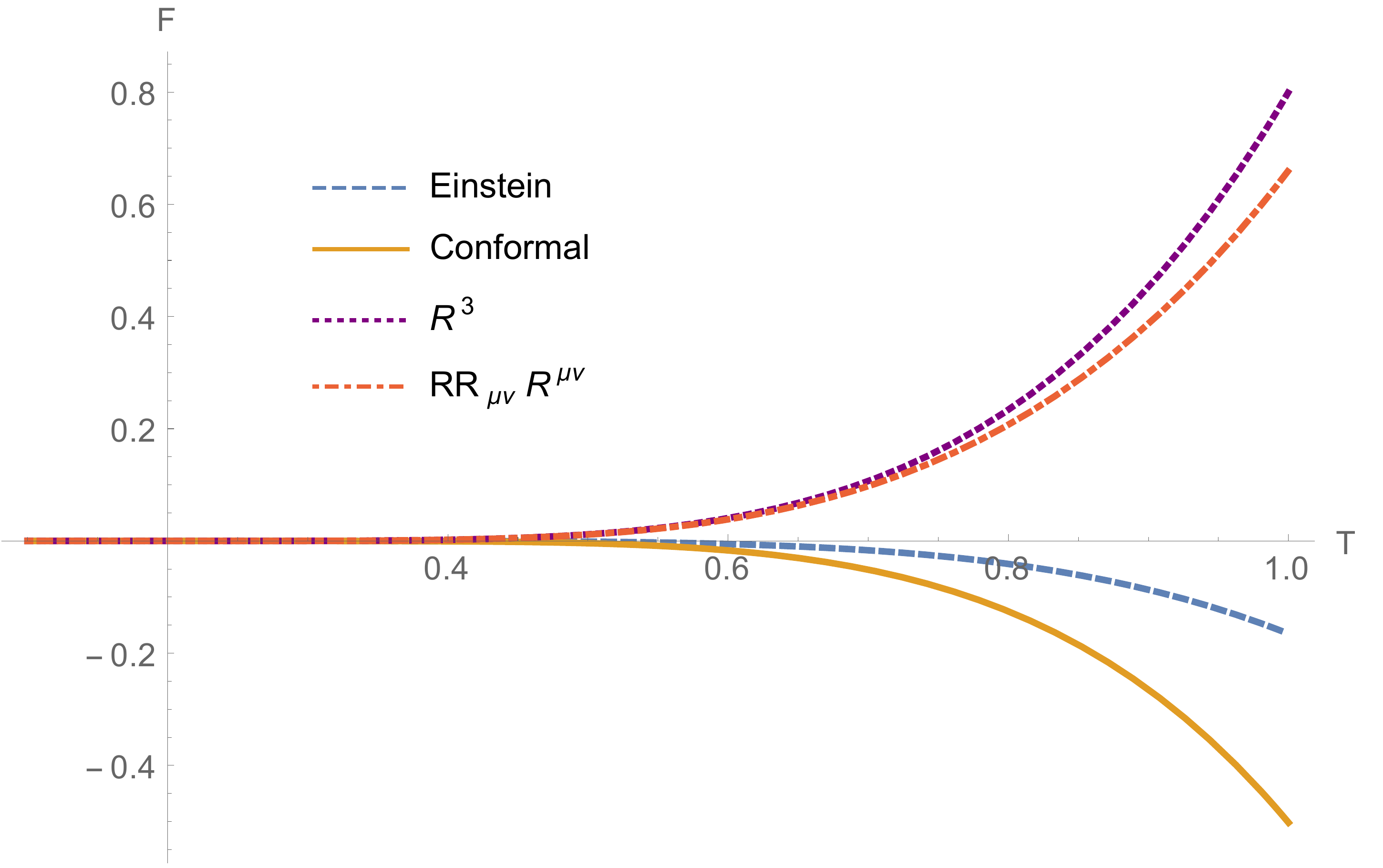}
	\caption{One-loop correction to the free energy in four special cases of $f(R,R_{\mu\nu}R^{\mu\nu})$. We have assumed $\Lambda=-1$.}
	\label{fig11}
\end{figure}

\begin{figure}
	\centering
	\includegraphics[scale=0.35]{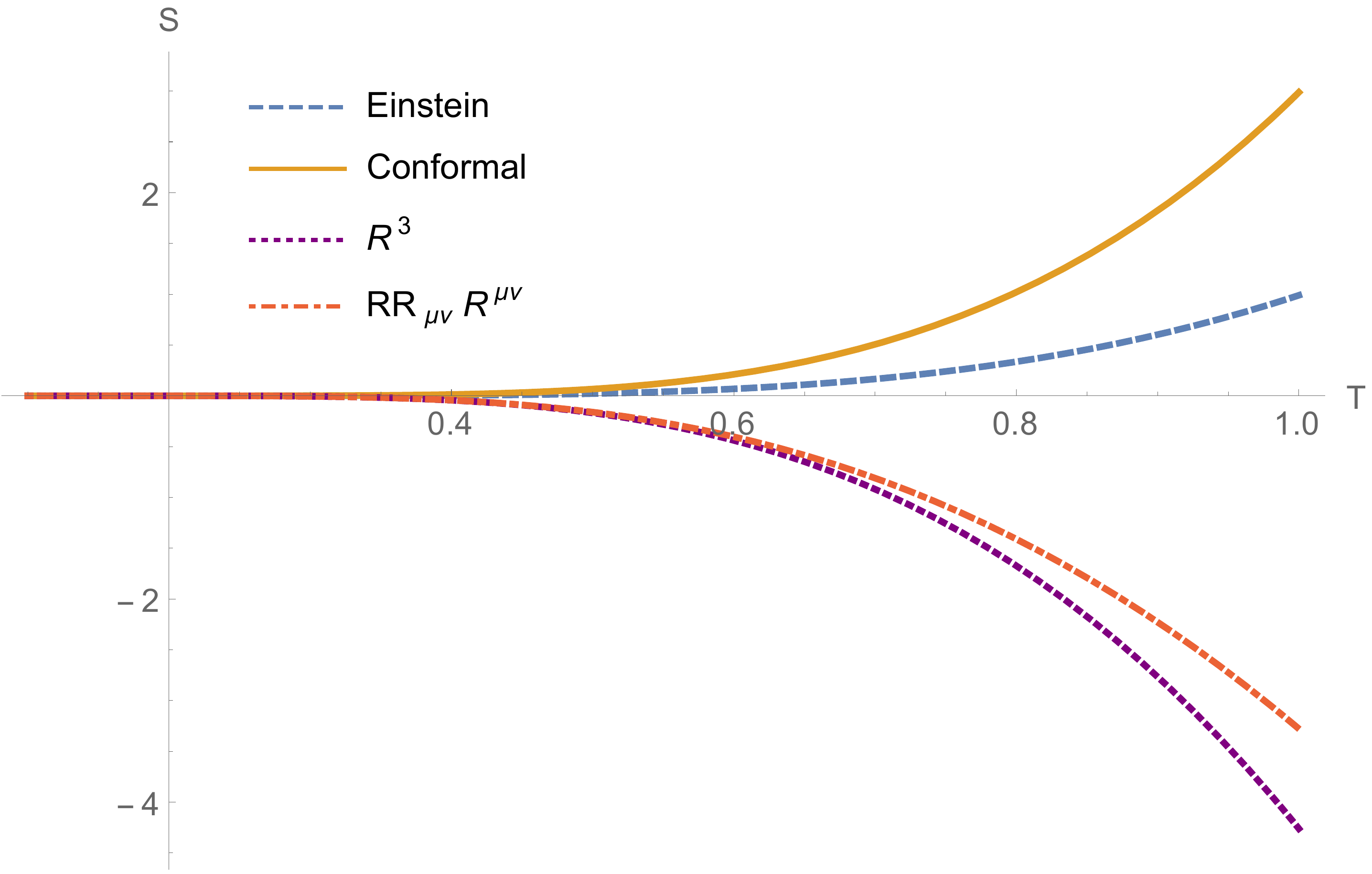}
	\caption{One-loop correction to the entropy in four special cases of $f(R,R_{\mu\nu}R^{\mu\nu})$. We have assumed $\Lambda=-1$.}
	\label{fig2}
\end{figure}

Using equation \eqref{pf1}, one can obtain the partition function of the general $f(R,R_{\mu\nu}R^{\mu\nu})$ gravity theory in a thermal quotient of AdS$_4$. In this section, we consider thermodynamical relations related to the thermal partition function \eqref{pf1} in four special cases discussed in section \ref{secpf}. Notice that the one-loop correction to the Helmhotz free energy can be obtained from the one-loop partition function via
$$F^{\textmd{1-loop}}=-\f1\beta\ln Z^{\textmd{1-loop}}.$$
In the case of the Einstein gravity with Lagrangian $\mathcal{L}=\kappa^2(R-2\Lambda)$ one can obtain the free energy as
\begin{align}
F_{\textmd{Einstein}}^{\textmd{1-loop}}=-\frac{1}{2}\sum_{k}\frac{e^{-\frac{3}{2}k\beta}}{\left(1-e^{-k\beta}\right)^3k\beta}\bigg[5e^{-k\beta\sqrt{\frac{17}{4}+\frac{2\Lambda}{3}}}-3e^{-k\beta\sqrt{\frac{13}{4}-\Lambda}}\bigg].
\end{align}
In the case of the conformal gravity with $\mathcal{L}=\f23\eta(3R_{\mu\nu}R^{\mu\nu}-R^2)$
we obtain
\begin{align}
F_{\textmd{Conformal}}^{\textmd{1-loop}}&=-\frac{1}{2}\sum_{k}\frac{e^{-\frac{3}{2}k\beta}}{\left(1-e^{-k\beta}\right)^3k\beta}\nonumber\\ &\times\bigg[5e^{-k\beta\sqrt{\frac{17}{4}+\frac{2\Lambda}{3}}}-3e^{-k\beta\sqrt{\frac{13}{4}-\Lambda}}-e^{-k\beta\sqrt{\frac{9}{4}-\frac{4\Lambda}{3}}}+5e^{-k\beta\sqrt{\frac{17}{4}+\frac{4\Lambda}{3}}}\bigg].
\end{align}
One can see that the first two terms in the second line correspond to the Einstein's theory.
For the Lagrangian $\mathcal{L}=\kappa^2 R+\eta R^3$, the free energy can be obtained as
\begin{align}
F_{\kappa^2 R+\eta R^3}^{\textmd{1-loop}}&=-\frac{1}{2}\sum_{k}\frac{e^{-\frac{3}{2}k\beta}}{\left(1-e^{-k\beta}\right)^3k\beta}\nonumber\\ &\times\bigg[5e^{-k\beta\sqrt{\frac{17}{4}+\frac{2\Lambda}{3}}}-3e^{-k\beta\sqrt{\frac{13}{4}-\Lambda}}-6e^{-k\beta\sqrt{\frac{13}{4}+2\Lambda}}-e^{-k\beta\sqrt{\frac{9}{4}-\frac{4\Lambda}{9}}}\bigg].
\end{align}
The last example is $\mathcal{L}=\kappa^2R+\eta RR_{\mu\nu}R^{\mu\nu}$ with the free energy
\begin{align}
&F_{\kappa^2R+\eta RR_{\mu\nu}R^{\mu\nu}}^{\textmd{1-loop}}=-\frac{1}{2}\sum_{k}\frac{e^{-\frac{3}{2}k\beta}}{\left(1-e^{-k\beta}\right)^3k\beta}\nonumber\\ &\qquad\quad \times\bigg[5e^{-k\beta\sqrt{\frac{17}{4}+\frac{2\Lambda}{3}}}-3e^{-k\beta\sqrt{\frac{13}{4}-\Lambda}}-6e^{-k\beta\sqrt{\frac{13}{4}+2\Lambda}}-e^{-k\beta\sqrt{\frac{9}{4}-\frac{2\Lambda}{5}}}+5e^{-k\beta\sqrt{\frac{17}{4}-\frac{10\Lambda}{3}}}\bigg].
\end{align}
Also, one can obtain the one-loop quantum correction to the entropy as
\begin{align}\label{ent}
S^{\textmd{1-loop}}=-\frac{\partial F^{\textmd{1-loop}}}{\partial T},
\end{align}
where $T=k\beta$ is the temperature.
In figures \ref{fig11} and \ref{fig2}, we have plotted the above expressions for the one-loop correction to the free energy as well as the entropy, for all the sample theories. 
One can see from the figures that the correction of the free energy in the case of the Einstein's gravity and conformal gravity is negative, indicating that the quantum correction decreases the background values of free energy. 
In other two cases, where the Lagrangian has a third order term in the curvature tensor, the quantum correction to the free energy increases the value of the classical free energy. 
This can be described by the fact that higher order curvature terms dominate the Einstein's gravity in the microscopic level. 
The quantum correction to the entropy is very similar to the effects of the free energy, however in a reverse direction due to the minus sign in \eqref{ent}.
%===================================================
\section{Conclusions and final remarks}
In this paper we have computed the partition function at the one-loop level for a modified theory of gravity containing a function of both the Ricci scalar and the square of the Ricci tensor.
Studying this theory can be very informative in the context of quantum gravity, since it can be thought of as a generalization of many other important modified gravity theories.
For our purpose we have used the usual Faddeev-Popov and De Witt method to find the total partition function at the one-loop level over a constant curvature background metric.
Then, using the York decomposition we were able to calculate this partition function in terms of the determinants of scalar, vector and tensor modes of the metric tensor.
We have moreover showed that the theory chosen here has \ads solution by presenting an example in the paper. 
Therefore, we were able to find the one-loop partition function of our theory over the \ads background, which is of great interest in other contexts, for instance in the AdS/CFT correspondance.
In this regard, we have also presented the calculated one-loop partition function over the \ads solution for 4 different gravity theories at the end of section \ref{secpf}.
Then, we have used the heat kernel method to compute the determinants of different operators in the partition function. 
The heat kernel representation is interesting since it shows a way to construct a traced heat kernel of the AdS$_4$ background. 
We have used this procedure to compute the one-loop thermal partition function of the theory over the AdS$_4$ space. 
This should be considered as a one-loop correction to the background partition function of the dual CFT theory. 
We have then straightforwardly computed the one-loop quantum corrections to both the Helmholtz free energy and the entropy of the theories we are interested in. 
It is interesting that for both the Einstein's theory and the conformal gravity theory these corrections increase the entropy from its value at the background, while for third order modified gravities such as $R^3$ or $RR_{\mu\nu}R^{\mu\nu}$ theory they decrease the background entropy from its background value. 
This can be traced back to the fact that the contribution coming from the higher order terms in the curvature tensor dominate that of the Einstein's gravity at small scales and hence compensate the effect of Einstein's gravity theory.
%===================================================
\appendix
\section{Operators in the second order variation of the action}\label{app1}
\begin{align}
{\cal{O}}_{_{\!\!AA}}=\fX\,\Box^2+\left(\fR+\frac{2\Lambda}{3}\fX\right)\Box
+\left( f-\frac{4}{9}\Lambda\left( 6\fR+11\Lambda \fX\right)\right).
\end{align}
\begin{align}
{\cal{O}}_{_{\!\!TT}}=2c_1(\Box+\Lambda).
\end{align}
\begin{align}
{\cal{O}}_{_{\!\!\chi\chi}}=&\f32(2\fX+3c_2)\Box^4-\f12\bigg(3\fR-8\Lambda(\fX+3c_2)\bigg)\Box^3\nonumber\\
&+\bigg(3f-8\Lambda\fR+4\Lambda^2(2c_2-3\fX)\bigg)\Box^2-4\Lambda c_1\Box.
\end{align}
\begin{align}
{\cal{O}}_{_{\!\!hh}}=\f{3}{8}(2\fX+3c_2)\Box^2-\f{1}{8}\bigg(3\fR-8\Lambda(\fX+3c_2)\bigg)\Box+\f14\bigg(f-4\Lambda\fR+4\Lambda^2(2c_2-\fX)\bigg).
\end{align}
\begin{align}
{\cal{O}}_{_{\!\!h\chi}}=a_3\Box^3+a_2\Box^2+a_1\Box,
\end{align}
where
\begin{align}
&a_1=2\Lambda(\fR-4\Lambda c_2),\\
&a_2=\f12\left[3\fR-8\Lambda(\fX+3c_2)\right],\\
&a_3=-\f32(2\fX+3c_2),
\end{align}
and
\begin{align}
c_1&=2\Lambda(\fR+2\Lambda\fX)-f,\\
c_2&=\fRR+4\Lambda(\fXR+\Lambda\fXX).
\end{align}
One can easily check that the following relation holds
\begin{align}
\f{3}{4\Lambda}a_1-a_2+\f{4\Lambda}{3}a_3=0.
\end{align}
%------------------------------------------------------------
\section{The masses}\label{app2}
\begin{align}
m^2_{A,\pm}&=\f{1}{6f_X}\left(-3f_R-2\Lambda f_X\pm3\sqrt{f_R^2-4f\,f_X+12\Lambda f_Rf_X+20\Lambda^2 f_X^2}\right),\\
m^2_{T,\pm}&=\f{1}{2}\left(\Lambda \pm\sqrt{9\Lambda^2+16\Lambda f_R-8f+32\Lambda^2 f_X}\right).
\end{align}
\begin{align}
m^2_{h,\pm}&=\left\{12
	f_{X}+8 \left[2 \rho ^2 \gamma+9 \Lambda  (f_{XX}
	\Lambda +f_{XR})\right]+18 f_{RR}\right\}^{-1}\nonumber\\&\times\bigg\{\bigg[8 (4		\Lambda  (\Lambda  (f_{X}-2 (4 \Lambda  (f_{XX} \Lambda
		+f_{XR})+f_{RR}))+f_R)-f)\nonumber\\&\times \left(6 f_{X}+4 \left(2
		\rho ^2 \gamma +9 \Lambda  (f_{XX} \Lambda
		+f_{XR})\right)+9 f_{RR}\right)\nonumber\\&+\left(3 f_R-8 \Lambda 
		\left(-\rho ^2+f_{X}+12 \Lambda  (f_{XX} \Lambda +f_{XR})+3
		f_{RR}\right)\right)^2\bigg]^{1/2}\nonumber\\&-8 \Lambda  \left(-\rho^2+f_{X}+12 \Lambda
	(f_{XX} \Lambda +f_{XR})+3 f_{RR}\right)+3 f_R\bigg\}.
\end{align}

\begin{align}
m^2_{\chi,\pm}&=\left\{8 f_{X}+6 [\gamma+8 \Lambda  (f_{XX} \Lambda +f_{XR})+2 f_{RR}]\right\}^{-1}\nonumber\\&\times\bigg\{-4 \gamma \Lambda +\bigg[\left(\Lambda  \left(4 \gamma+32
		f_{XX} \Lambda ^2+32 f_{XR} \Lambda +8 f_{RR}-3\right)-2
	f_R\right)^2\nonumber\\&-16 (f-\Lambda  (4 f_{X} \Lambda +2
		f_R+\Lambda )) (4 f_{X}+3 (\gamma+8 \Lambda  (f_{XX}
		\Lambda +f_{XR})+2 f_{RR}))\bigg]^{1/2}\nonumber\\&-32 f_{XX} \Lambda ^3+2 f_R-32
	f_{XR} \Lambda ^2-8 f_{RR} \Lambda +3 \Lambda \bigg\}.
\end{align}
\section{The decomposition determinants}\label{app3}
In this section we compute $Z_{N_{h}}$ which is appeared in the expression of the partition function. 
The determinants $Z_{N_{C}}$ and $Z_{N_{b}}$ can be obtained in a similar way. First note that
\begin{align}\label{ap1}
	1=\int Dh_{\mu\nu}e^{-\int d^4x\sqrt{-g}h_{\mu\nu}h^{\mu\nu}}.
\end{align}
We want to compute the Jacobian of the transformation which is in the form
\begin{align}
Dh_{\mu\nu}=Z_{N_h}DA_{\mu\nu}DT_\mu D\chi Dh.
\end{align}
Inserting the relation \eqref{317} into \eqref{ap1} and integrating by parts, one can obtain
\begin{align}
1&=\int Z_{N_h}DA_{\mu\nu}DT_\mu D\chi Dh \,\nonumber\\&\times\textmd{exp}\left\{-\int d^4x\sqrt{-g}\left[A_{\mu\nu}A^{\mu\nu}+\f14h^2+2T_\mu(-\Box-\Lambda)T^\mu+3\chi(-\Box)\left(-\Box-\f{4\Lambda}{3}\right)\chi\right]\right\}.
\end{align}
Using the Gaussian integral for the fields $A_{\mu\nu}$, $T_\mu$, $h$ and $\chi$, one can obtain $Z_{N_h}$ as
\begin{align}
Z_{N_h}=\left[\det_{(1)}(-\Box-\Lambda)\det_{(0)}(-\Box)\det_{(0)}\left(-\Box-\f{4\Lambda}{3}\right)\right]^{1/2}.
\end{align}
Similarly, one can find
\begin{align}
Z_{N_C}={\det_{(0)}}^{-1}(-\Box),\qquad Z_{N_b}=\sqrt{\det_{(0)}(-\Box)}.
\end{align}


\begin{thebibliography}{99}
	\bibitem{darkenergy} A. De Felice and S. Tsujikawa, Living Rev. Rel. {\bf 13}, 3 (2010); T. P. Sotiriou and V. Faraoni, Rev. Mod. Phys. {\bf 82}, 451 (2010); S. Nojiri and S. D. Odintsov, Phys. Rept. {\bf 505}, 59 (2011), arXiv:1011.0544; S. Nojiri, S. D. Odintsov, and V. K. Oikonomou, Phys. Rept. {\bf 692}, 1 (2017), arXiv:1705.11098.
	\bibitem{nongrav} G. 't Hooft and M. Veltman, Ann. Inst. Henri Poincare
	{\bf 20}, 69 (1974); S. Deser and P. van Nieuwenhuizen, Phys.
	Rev. D {\bf 10}, 401 (1974); {\bf 10}, 411 (1974); S. Deser, P. van
	Nieuwenhuizen, and H. S. Tsao, Phys.
	Rev. D {\bf 10}, 3337 (1974).
	\bibitem{renor} B. Utiyama and B. DeWitt, J. Math. Phys. {\bf 3}, 608
	(1962); S. Deser, in Gauge Theories and Modem Field
	Theory, proceedings of the Boston Conference, 1975,
	edited by H. Axnowitt and P. Nath (MIT, Cambridge,
	Mass., 1976).
	\bibitem{quantfR} M. S. Ruf and C. F. Steinwachs, Phys. Rev. D {\bf 97}, 044049 (2018); M. S. Ruf and C. F. Steinwachs, Phys. Rev. D {\bf 97}, 044050 (2018); G. Cognola, E. Elizalde, S. Nojiri, S. D. Odintsov, and S. Zerbini,  	JCAP {\bf 02}, 010 (2005); G. Cognola and S. Zerbini, J. Phys. A: Math. Theor. {\bf 45}, 374014 (2012), arXiv:1203.5032 [gr-qc].
	
	\bibitem{horava} P. Horava, Phys. Rev. D {\bf 79}, 084008 (2009). 
	\bibitem{horavaapp} E. Kiritsis and G. Kofinas, Nucl. Phys. B{\bf821}, 467 (2009); P. Horava, Phys. Rev. Lett. {\bf 102}, 161301 (2009); G. Amelino-Camelia, Living Rev. Rel. {\bf 16}, 5 (2013); D. Orlando and S. Reffert, Phys. Rev. D {\bf 80}, 041501 (2009);E. N. Saridakis, Eur. Phys. J. C {\bf 67}, 229 (2010). 
	\bibitem{lorenthorava} Z. Haghani, T. Harko, H. R. Sepangi, and S. Shahidi, arXiv:1404.7689 [gr-qc]; T. Jacobson and D. Mattingly, Phys. Rev. D {\bf 64}, 024028 (2001); D. Blas, O. Pujol`as, and S. Sibiryakov, JHEP {\bf 10}, 029 (2009).
	\bibitem{stelle} K. Stelle, Phys. Rev. D {\bf 16}, 953 (1977).
	\bibitem{dsfR} K. Bamba, G. Cognola, S. D. Odintsov, and S. Zerbini, Phys. Rev. D {\bf 90}, 023525 (2014), arXiv:1404.4311 [gr-qc].
	
	\bibitem{maldacena}J. M. Maldacena, Adv. Theor. Math. Phys. {\bf 2}, 231 (1998).
	\bibitem{odintsov} I. L. Buchbinder, S. Odintsov, and L. Shapiro, Effective Action in Quantum Gravity, CRC Press, 1992.
	\bibitem{iva} R. Gopakumar, R. Kumar Gupta, and S. Lal, JHEP {\bf 11}, 010 (2011); J. R. David, M. R. Gaberdiel, and R. Gopakumar, JHEP {\bf 04},125 (2010); M. Irakleidou and I. Lovrekovic, Phys. Rev. D {\bf 93}, 104043 (2016), arXiv:1512.07130 [hep-th].
	\bibitem{akhary}N. Ohta, R. Percacci, and A. D. Pereira, arXiv:1804.01608 [hep-th].
	\bibitem{tseytlin} E. S. Fradkin and A. A. Tseytlin, Nucl. Phys. B{\bf234}, 472 (1984).
\end{thebibliography}
\end{document}